\documentclass{MICE}
\usepackage{lineno}
\usepackage{times}
\usepackage{bm}         
\usepackage{amsmath,amssymb,bm,slashed} 
\usepackage{amssymb}    
\usepackage{graphicx}   
\usepackage{verbatim}   
\usepackage{color}      
\usepackage{subfigure}  
\usepackage{hyperref}   
\usepackage{enumerate}

\usepackage{fmtcount}   

\usepackage[square,comma,numbers,sort&compress]{natbib}

\pdfoutput=1

\begin{document}

\thispagestyle{empty}

\begin{tabular}{p{0.175\textwidth} p{0.5\textwidth} p{0.225\textwidth}}
  \hspace{-0.8cm}\leftline{\today}                      &
  \centering{ Muon Ionization Cooling Experiment}                  &
  \rightline{RAL-P-2017-002} 
\end{tabular}
\vspace{-1.0cm}\\
\rule{\textwidth}{0.43pt}

\begin{center}
  {\bf
    {\LARGE Design and expected performance of the MICE demonstration of ionization cooling} \\
  }
  \vspace{0.2cm}
  The MICE collaboration \\
  \vspace{-0.0cm}
\end{center}

\makeatletter

\newcommand{\bra}[1]{\ensuremath{\langle #1 |}}   
\newcommand{\ket}[1]{\ensuremath{| #1 \rangle}}   
\newcommand{\bigbra}[1]{\ensuremath{\big\langle #1 \big|}}   
\newcommand{\bigket}[1]{\ensuremath{\big| #1 \big\rangle}}   
\newcommand{\amp}[3]{\ensuremath{\left\langle #1 \,\left|\, #2%
                     \,\right|\, #3 \right\rangle}}  
\newcommand{\sprod}[2]{\ensuremath{\left\langle #1 |%
                     #2 \right\rangle}}  
\newcommand{\ev}[1]{\ensuremath{\left\langle #1 %
                     \right\rangle}} 
\newcommand{\ds}[1]{\ensuremath{\! \frac{d^3#1}{(2\pi)^3 %
                     \sqrt{2 E_\vec{#1}}} \,}} 
\newcommand{\dst}[1]{\ensuremath{\! %
                     \frac{d^4#1}{(2\pi)^4} \,}} 
\newcommand{\tr}{\text{tr}}
\newcommand{\sgn}{\text{sgn}}
\newcommand{\diag}{\text{diag}}
\newcommand{\BR}{\text{BR}}

\renewcommand{\vec}[1]{{\mathbf{#1}}}
\renewcommand{\Re}{{\text{Re}}}
\renewcommand{\Im}{{\text{Im}}}
\newcommand{\iso}[2]{{\ensuremath{{}^{#2}}\ensuremath{\rm #1}}}
\newcommand{\eps}{{\ensuremath{\epsilon}}}
\newcommand{\draftnote}[1]{{\bf\color{red} \MakeUppercase{#1}}}
\newcommand{\panm}[1]{{\color{blue} #1}}
\providecommand{\abs}[1]{\lvert#1\rvert}
\providecommand{\norm}[1]{\lVert#1\rVert}
\newcommand{\gsim}      {\mbox{\raisebox{-0.4ex}{$\;\stackrel{>}{\scriptstyle \sim}\;$}}}
\newcommand{\lsim}      {\mbox{\raisebox{-0.4ex}{$\;\stackrel{<}{\scriptstyle \sim}\;$}}}

\def\parenbar{\mathpalette\p@renb@r}
\def\p@renb@r#1#2{\vbox{%
  \ifx#1\scriptscriptstyle \dimen@.7em\dimen@ii.2em\else
  \ifx#1\scriptstyle \dimen@.8em\dimen@ii.25em\else
  \dimen@1em\dimen@ii.4em\fi\fi \offinterlineskip
  \ialign{\hfill##\hfill\cr
    \vbox{\hrule width\dimen@ii}\cr
    \noalign{\vskip-.3ex}%
    \hbox to\dimen@{$\mathchar300\hfil\mathchar301$}\cr
    \noalign{\vskip-.3ex}%
    $#1#2$\cr}}}

%
\providecommand{\anmne}{\mbox{$\bar\nu_{\mu} \rightarrow \bar\nu_e$}} 
\providecommand{\nmne}{\mbox{$\nu_{\mu}\rightarrow\nu_e$}} 
\providecommand{\anm}{\mbox{$\bar\nu_\mu$}} 
\providecommand{\nm}{\mbox{$\nu_\mu$}}
\providecommand{\nue}{\mbox{$\nu_e$}} 
\providecommand{\ane}{\mbox{$\bar\nu_e$}} 
\providecommand{\enu}{\mbox{$E_\nu$}}
\providecommand{\piz}{\mbox{$\pi^0 $}}
\providecommand{\pip}{\mbox{$\pi^+$}} 
\providecommand{\pim}{\mbox{$\pi^-$}}

\parindent 10pt
\pagestyle{plain}
\pagenumbering{arabic}                   
\setcounter{page}{1}

\begin{quotation}

\noindent
Muon beams of low emittance provide the basis for the intense,
well-characterised neutrino beams necessary to elucidate the physics
of flavour at a neutrino factory and to provide lepton-antilept\-on
collisions at energies of up to several TeV at a muon collider. 
The international Muon Ionization Cooling Experiment (MICE) aims to
demonstrate ionization cooling, the technique by which it is proposed
to reduce the phase-space volume occupied by the muon beam at such
facilities. 
In an ionization-cooling channel, the muon beam passes through a
material in which it loses energy. 
The energy lost is then replaced using RF cavities. 
The combined effect of energy loss and re-acceleration is to reduce
the transverse emittance of the beam (transverse cooling). 
A major revision of the scope of the project was carried out over the
summer of 2014. 
The revised experiment can deliver a demonstration of ionization
cooling.
The design of the cooling demonstration experiment will be described
together with its predicted cooling performance.

\end{quotation}

\section{Introduction}
\label{Sect:Intro}

Stored muon beams have been proposed as the source of neutrinos at a
neutrino factory \cite{Geer:1997iz,Apollonio:2002en} and as the means
to deliver multi-TeV lepton-antilepton collisions at a muon collider
\cite{Neuffer:1994bt,Palmer:2014nza}.
In such facilities the muon beam is produced from the decay of pions
generated by a high-power proton beam striking a target. 
The tertiary muon beam occupies a large volume in phase space. 
To optimise the muon yield while maintaining a suitably small aperture
in the muon-acceleration system requires that the muon beam be
``cooled'' (i.e., its phase-space volume reduced) prior to
acceleration.
A muon is short-lived, decaying with a lifetime of 2.2\,$\mu$s in its
rest frame.
Therefore, beam manipulation at low energy ($\lsim 1$\,GeV)
must be carried out rapidly.
Four cooling techniques are in use at particle accelerators:
synchrotron-radiation cooling \cite{2012acph.book.....L}; laser
cooling 
\cite{PhysRevLett.64.2901,PhysRevLett.67.1238,doi:10.1063/1.329218}; 
stochastic cooling \cite{Marriner:2003mn}; and electron cooling
\cite{1063-7869-43-5-R01}.
Synchrotron-radiation cooling is observed only in electron or positron
beams, owing to the relatively low mass of the electron. 
Laser cooling is limited to certain ions and atomic beams.
Stochastic cooling times are dependent on the bandwidth of the
stochastic-cooling system relative to the frequency spread of the
particle beam. 
The electron-cooling time is limited by the available electron density
and the electron-beam energy and emittance. 
Typical cooling times are between seconds and hours, long compared
with the muon lifetime. 
Ionization cooling proceeds by passing a muon beam through a material,
the absorber, in which it loses energy through ionization, and
subsequently restoring the lost energy in accelerating cavities.
Transverse and longitudinal momentum are lost in equal proportions in
the absorber, while the cavities restore only the momentum component
parallel to the beam axis.
The net effect of the energy-loss/re-acceleration process is to
decrease the ratio of transverse to longitudinal momentum, thereby
decreasing the transverse emittance of the beam.
In an ionization-cooling channel the cooling time is short enough to
allow the muon beam to be cooled efficiently with modest decay
losses.
Ionization cooling is therefore the technique by which it is proposed
to cool muon beams \cite{cooling_methods,Neuffer:1983xya,Neuffer:1983jr}.
This technique has never been demonstrated experimentally and such a
demonstration is essential for the development of future
high-brightness muon accelerators. 

The international Muon Ionization Cooling Experiment (MICE)
collaboration proposes a two-part process to perform a full
demonstration of transverse ionization cooling.   
First, the ``Step~IV'' configuration \cite{Rajaram:2015bra} will be
used to study the material and beam properties that determine the
performance of an ionization-cooling lattice.  
Secondly, a study of transverse-emittance reduction in a cooling cell
that includes accelerating cavities will be performed. 

The cooling performance of an ionization-cooling cell depends on the
emittance and momentum of the initial beam, on the properties of the
absorber material and on the transverse betatron function
($\beta_{\perp}$) at the absorber.  
These factors will be studied using the  Step IV configuration.  
Once this has been done, ``sustainable'' ionization cooling must be
demonstrated.   
This requires restoring energy lost by the muons as they pass through
the absorber using RF cavities.  
The experimental configuration with which the MICE collaboration
originally proposed to study ionization cooling was presented
in~\cite{MICEproposal:2003}.  
This configuration was revised to accelerate the timetable on which a
demonstration of ionization cooling could be delivered and to reduce
cost.
This paper describes the revised lattice  proposed by the MICE
collaboration for the demonstration of ionization cooling and presents
its performance.

\graphicspath{{02-Cooling/Figures/}}

\section{Cooling in neutrino factories and muon colliders}
\label{Sect:cooling}

At production, muons occupy a large volume of phase space.
The emittance of the initial muon beam must be reduced before the beam
is accelerated.
A neutrino factory \cite{Apollonio:2008aa} requires the transverse
emittance to be reduced from 15--20\,mm to 2--5\,mm. 
A muon collider \cite{Ankenbrandt:1999cta} requires the muon beam to
be cooled in all six phase-space dimensions; to achieve the desired
luminosity requires an emittance of $\sim 0.025$\,mm in the transverse
plane and $\sim 70$\,mm in the longitudinal direction 
\cite{Alsharoa:2002wu,Palmer:2007zzc}.

Ionization cooling is achieved by passing a muon beam through a 
material with low atomic number ($Z$), in which it loses energy by
ionization, and subsequently accelerating the beam.
The rate of change of  the normalised transverse emittance,
$\varepsilon_\perp$, is given approximately by 
\cite{Neuffer:1983xya,Penn:2000mt,Rogers08beamdynamics}:
\begin{equation}
  \frac{\text{d}\varepsilon_\perp}{\text{d}z}\backsimeq
  -\frac{\varepsilon_\perp}{\beta^2E_{\mu}}\Big\langle
  \frac{\text{d}E}{\text{d}z} \Big\rangle +
  \frac{\beta_{\perp}(13.6\,\text{MeV/}c)^2}{2\beta^3E_{\mu}m_{\mu}X_0} \, ;
  \label{cool_eq}
\end{equation}
where $z$ is the longitudinal coordinate, $\beta c$ is the muon
velocity, $E_{\mu}$ the energy,
$\big\langle\frac{\text{d}E}{\text{d}z}\big\rangle$ the mean rate of
energy loss per unit path-length, $m_{\mu}$ the mass of the muon, $X_0$
the radiation length of the absorber and $\beta_{\perp}$ the
transverse betatron function at the absorber. 
The first term of this equation describes ``cooling'' by ionization
energy loss and the second describes ``heating'' by multiple Coulomb
scattering.
Equation \ref{cool_eq} implies that the equilibrium emittance, for
which $\frac{\text{d}\varepsilon_\perp}{\text{d}z}=0$, and the
asymptotic value of $\frac{\text{d}\varepsilon_\perp}{\text{d}z}$ for
large emittance are functions of muon-beam energy.

In order to have good performance in an ionization-cooling channel,
$\beta_{\perp}$ needs to be minimised and  $X_0\langle
\frac{\text{d}E}{\text{d}z} \rangle$ maximised. 
The betatron function at the absorber is minimised using a suitable
magnetic focusing channel (typically
solenoidal) \cite{Stratakis:2014nna,Neuffer:2016gtw} and
$X_0\langle \frac{\text{d}E}{\text{d}z} \rangle$ is maximised using a 
low-$Z$ absorber such as liquid hydrogen (LH$_2$) or lithium hydride
(LiH) \cite{Tollestrup:2000zz}.

\graphicspath{{03-MICE_overview/Figures/}}

\section{The Muon Ionization Cooling Experiment}
\label{Sect:overview}

The muons for MICE come from the decay of pions produced at an
internal target dipping directly into the circulating proton beam in
the ISIS synchrotron at the Rutherford Appleton Laboratory
(RAL)~\cite{Booth:2012qz, Booth:2016lno}.  
A beam line of 9 quadrupoles, 2 dipoles and a superconducting ``decay
solenoid'' collects and transports the momentum-selected beam into the
experiment~\cite{Bogomilov:2012sr}.
The small fraction of pions that remain in the beam may be rejected
during analysis using the time-of-flight hodoscopes and Cherenkov
counters that are installed in the beam line upstream of the
experiment~\cite{Adams:2013lba}.
A diffuser is installed at the upstream end of the experiment to vary
the initial emittance of the beam. 
Ionization cooling depends on momentum through $\beta$, $E_{\mu}$ and 
$\Big\langle \frac{\text{d}E}{\text{d}z} \Big\rangle$ as shown in
equation \ref{cool_eq}.
It is therefore proposed that the performance of the cell be measured
for momenta in the range 140\,MeV/c to 240\,MeV/$c$
\cite{MICEproposal:2003}.

\subsection{The configuration of the ionization-cooling experiment}

\begin{figure}
  \begin{center}
    \includegraphics[width=\textwidth]{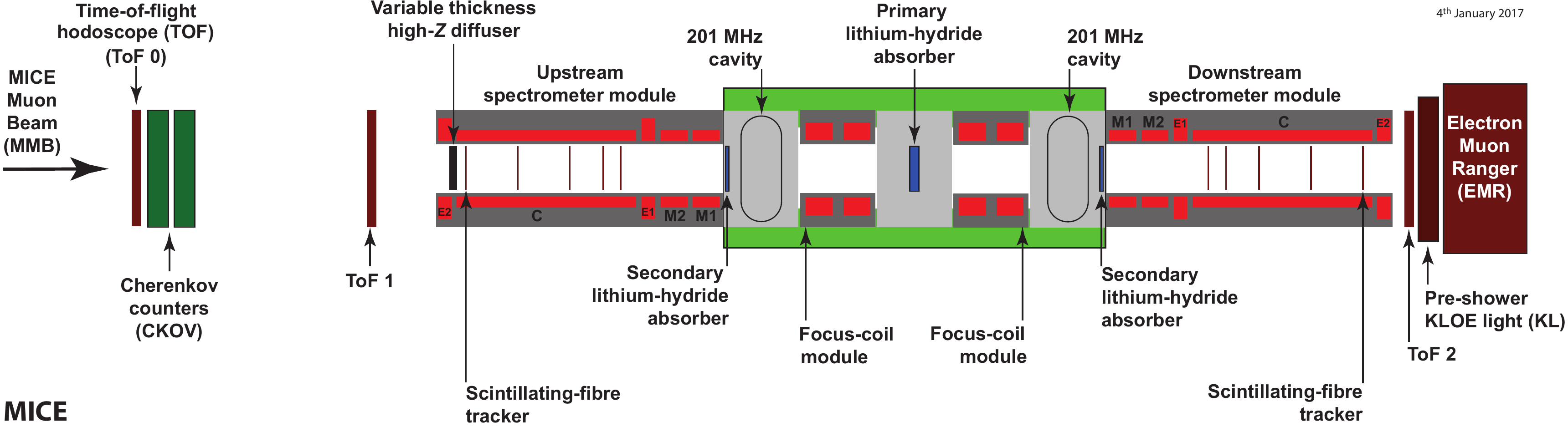}
  \end{center}
  \caption{
    Layout of the lattice configuration for the cooling demonstration.
    The red rectangles represent the solenoids. 
    The individual coils in the spectrometer solenoids are labelled
    E1, C, E2, M1 and M2. 
    The ovals represent the RF cavities and the blue rectangles the
    absorbers. 
    The various detectors (time-of-flight
    hodoscopes \cite{Bertoni:2010by,MICE:Note:286:2010}, Cerenkov
    counters \cite{Cremaldi:2009zj}, scintillating-fibre
    trackers \cite{Ellis:2010bb}, KLOE Light (KL)
    calorimeter \cite{Bogomilov:2012sr,Ambrosino2009239},
    electron muon ranger \cite{Asfandiyarov:2016erh}) used to
    characterise the beam are also represented. 
    The green-shaded box indicates the cooling
    cell.\label{Fig:Overview}
  }
\end{figure}

The configuration proposed for the demonstration of ionization cooling
is shown in figure \ref{Fig:Overview}.
It contains a cooling cell sandwiched between two
spectrometer-solenoid modules.
The cooling cell is composed of two 201\,MHz cavities, one primary
(65\,mm) and two secondary (32.5\,mm) LiH absorbers placed between
two superconducting ``focus-coil'' (FC) modules. 
Each FC has two separate windings that can be operated either with the
same or in opposed polarity.

The emittance is measured upstream and downstream of the cooling cell
using scintillating-fibre tracking detectors \cite{Ellis:2010bb}
immersed in the uniform 4\,T magnetic field provided by three
superconducting coils (E1, C, E2).
The trackers are used to reconstruct the trajectories of individual
muons at the entrance and exit of the cooling cell. 
The reconstructed tracks are combined with information from
instrumentation upstream and downstream of the spectrometer modules to
measure the muon-beam emittance at the upstream and downstream tracker
reference planes.
The instrumentation upstream and downstream of the spectrometer
modules serves to select a pure sample of muons.
Time-of-flight hodoscopes are used to determine the time at
which the muon crosses the RF cavities.
The spectrometer-solenoid magnets also contain two superconducting
``matching'' coils (M1, M2) that are used to match the optics between
the uniform field region and the neighbouring FC.

The secondary LiH absorbers (SAs) are introduced between the cavities
and the trackers to minimise the exposure of the trackers to
``dark-current'' electrons originating from the RF cavities.  
Experiments at the MuCool Test Area (MTA) at
Fermilab \cite{Leonova:2015mua} have observed that the rate of direct
X-ray production from the RF cavities can be managed to ensure it does
not damage the trackers \cite{Torun:2016aoe}. 
The SAs are introduced to minimise the exposure of the trackers to
energetic dark-current electrons that could produce background hits.
The SAs are positioned between the trackers and the cavities such that
they can be removed to study the empty channel.
The SAs increase the net transverse-cooling effect since the betatron
functions at these locations are small.

\begin{figure}
  \begin{center}
    \includegraphics[width=0.5\textwidth]{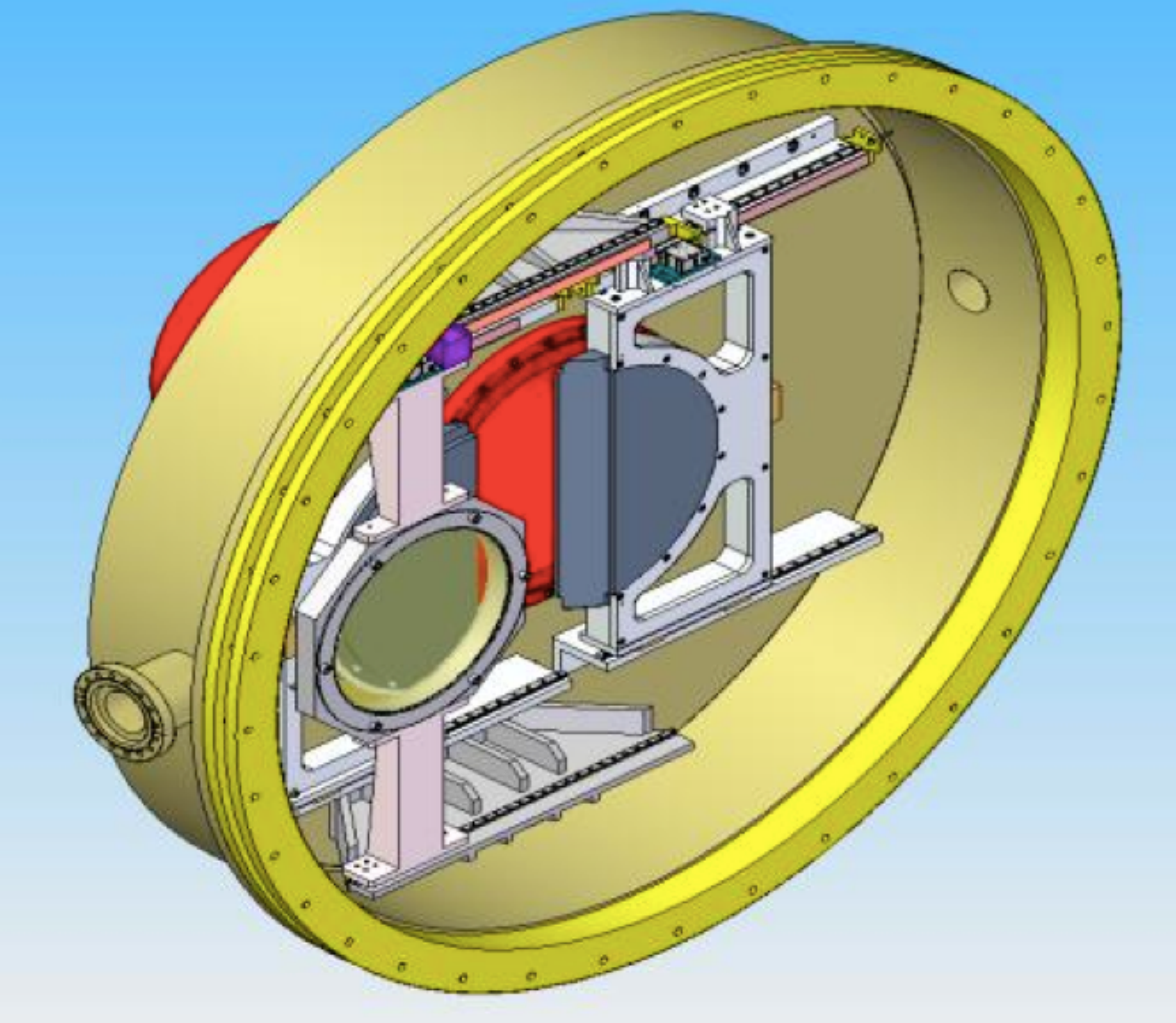}
  \end{center}
  \caption{
    Design of the movable frame for the secondary absorber (front) and
    the lead radiation shutter (back). 
    The half discs of the lead shutter (grey) can be seen together
    with the rails (white) inside the vacuum chamber (yellow).
  }
  \label{Fig:Shutters}
\end{figure}

Retractable lead radiation shutters will be installed on rails
between the spectrometer solenoids and the RF modules to protect the
trackers against dark-current induced radiation during cavity
conditioning. 
The SAs will be mounted on a rail system similar to that which will be
used for the lead shutters and will be located between the cavities
and the lead shutters. 
Both mechanisms will be moved using linear piezo-electric motors that
operate in vacuum and magnetic field. 
The design of both the radiation shutter and the movable SA inside the
vacuum chamber is shown in figure \ref{Fig:Shutters}. 

The RF cavities are 201\,MHz ``pillbox'' resonators, 430\,mm in
length, operating in the TM$_{010}$ mode with large diameter
apertures to accommodate the high emittance beam. 
The apertures are covered by thin (0.38\,mm) beryllium windows to
define the limits for the accelerating RF fields whilst minimising the
scattering of muons.
The cavity is excited by two magnetic-loop couplers on opposite sides
of the cavity.
At the particle rate expected in MICE there is no beam-loading of the
RF fields. 
An effective peak field of 10.3\,MV/m is expected for a drive power
of 1.6\,MW to each cavity.
This estimate was used to define the gradient in the simulations
described below.

\graphicspath{{04-Lattice_design/Figures/}}

\section{Lattice design}
\label{Sect:design}
\subsection{Design parameters}

The lattice has been optimised to maximise the reduction in transverse
emittance.
The optimum is obtained by matching the betatron function to a small
value in the central absorber while minimising its maximum values in
the FC modules; limiting the size of the betatron function in the
FCs helps to reduce the influence of non-linear terms in the
magnetic-field expansion.
The matching accounts for the change in energy of the muons as
they pass through the cooling cell by adjusting currents in the
upstream and downstream FCs and in the matching coils in the
spectrometer solenoids independently while maintaining the field in
the tracking volumes at 4\,T. 
In this configuration, it is also possible to keep the betatron
function relatively small at the position of the secondary absorbers
whilst maintaining an acceptable beam size at the position of the
cavities.

\begin{figure}
  \begin{center}
    \includegraphics[width=0.9\textwidth]{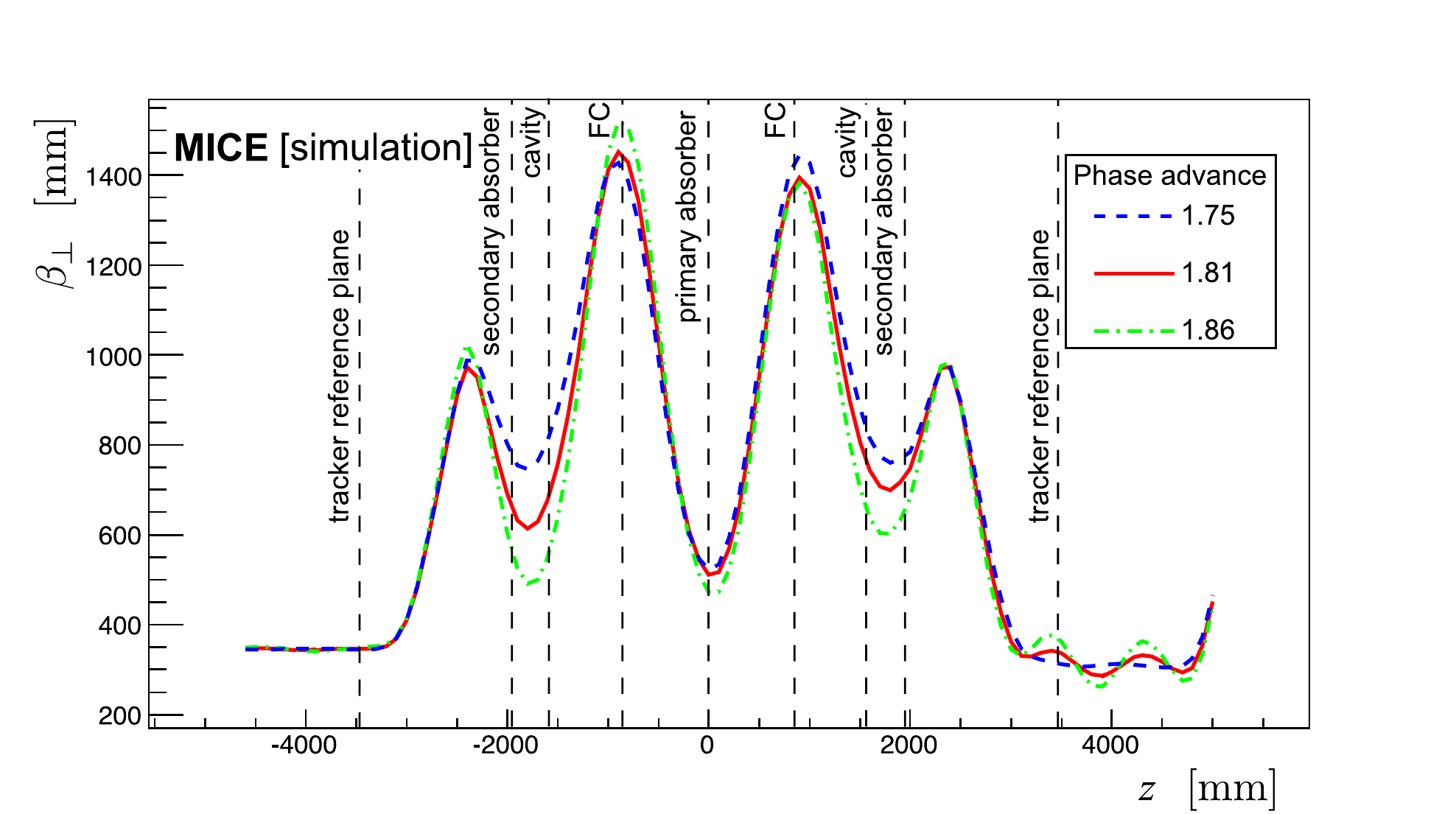}
  \end{center}
  \caption{
    Transverse 4D beta-function versus longitudinal coordinate $z$ in
    the cooling-demonstration lattice for 200\,MeV/$c$ settings with a
    phase advance of $2\pi\times1.75$ (dashed blue line),
    $2\pi\times1.81$ (solid red line) and $2\pi\times1.86$ (dot-dashed
    green line).
    The vertical dashed lines with labels show the positions of the
    tracker reference planes and the centres of the absorbers, RF
    cavities and focus coil modules. 
  }
  \label{Fig:BetaPhaseAdv}
\end{figure}

\begin{figure}
  \begin{center}
    \includegraphics[width=0.9\textwidth]{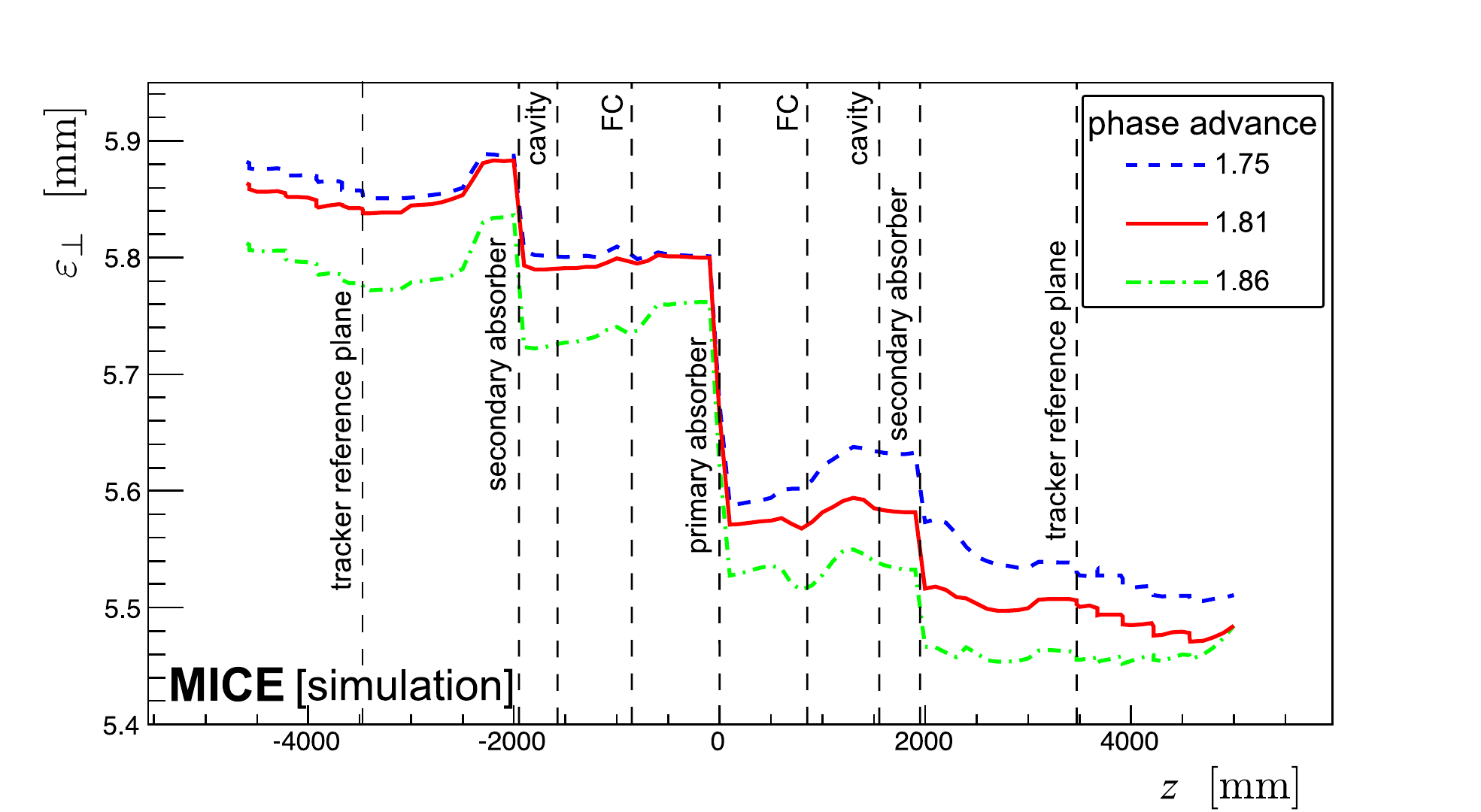}
  \end{center}
  \caption{
    4D emittance evolution in the cooling-demonstration lattice for
    200\,MeV/$c$ settings with a phase advance of $2\pi\times1.75$
    (dashed blue line), $2\pi\times1.81$ (solid red line) and
    $2\pi\times1.86$ (dot-dashed green line). 
    The vertical dashed lines with labels show the positions of the
    tracker reference planes and the centres of the absorbers, RF
    cavities and focus coil modules.
  }
  \label{Fig:EmitPhaseAdv}
\end{figure}

Chromatic aberrations caused by the large momentum spread of the beam
($\sim 5\%$ rms) lead to a chromatic mismatch of the beam in the
downstream solenoid unless the  phase advance across the cooling cell
(i.e., the rate of rotation of the phase-space ellipse) is chosen
appropriately.
The phase advance of the cell is obtained by integrating the inverse 
of the beta-function along the beam axis from the reference plane in
the upstream spectrometer-solenoid to the reference plane in the
downstream spectrometer-solenoid.
Such a mismatch reduces the effective transverse-emittance reduction
through the chromatic decoherence that results from the superposition
of beam evolutions for the different betatron frequencies that result
from the range of momenta in the beam.
For beams with a large input emittance, spherical aberrations may lead
to phase-space filamentation.
The chromatic and spherical aberrations were studied by tracking
samples of muons through the lattice using the ``MICE Analysis User
Software'' (MAUS, see section \ref{Sect:Simu}).
The betatron-function and emittance evolution of a 200\,MeV/$c$ beam
with the initial parameters given in table \ref{Tab:InitBeam} are
shown, for different phase advances, in figures \ref{Fig:BetaPhaseAdv}
and \ref{Fig:EmitPhaseAdv} respectively.
The phase advance of $2\pi\times1.81$ showed the largest
transverse-emittance reduction and was therefore chosen.
The lattice parameters for this phase advance are presented in table
\ref{Tab:LatticeParameters}.   
\begin{table}
  \caption{
    General parameters of the initial beam conditions used in the
    simulations.
  }
  \label{Tab:InitBeam}
  \begin{center}
    \begin{tabular}{lc}
      \hline
	Parameter & Value \\
      \hline
	Particle				&	muon $\mu^+$ \\
	Number of particles				&	10000 \\
	Longitudinal position [mm] 		& $-4612.1$\\
	Central energy (140\,MeV/$c$ settings) [MeV] & 175.4\\
	Central energy (200\,MeV/$c$ settings) [MeV] & 228.0\\
	Central energy (240\,MeV/$c$ settings) [MeV] & 262.2\\	
      \hline
	Transverse Gaussian distribution: & \\
      \hline
	$\alpha_{\perp}$ 				 & 0 \\
	$\beta_{\perp}$ (140\,MeV/$c$ settings) [mm] &233.5 \\
	$\varepsilon_{\perp}$ (140\,MeV/$c$ settings) [mm] &4.2 \\
	$\beta_{\perp}$ (200\,MeV/$c$ settings) [mm] & 339.0 \\
	$\varepsilon_{\perp}$ (200\,MeV/$c$ settings) [mm] &6.0 \\
	$\beta_{\perp}$ (240\,MeV/$c$ settings) [mm] & 400.3 \\
	$\varepsilon_{\perp}$ (240\,MeV/$c$ settings) [mm] &7.2 \\
      \hline
	Longitudinal Gaussian distribution: & \\
      \hline
	Longitudinal emittance [mm]		& 20\\
	Longitudinal $\beta$	[ns]			&	11\\
	Longitudinal $\alpha$			& $-0.7$\\	
	rms momentum spread (140\,MeV/$c$ settings)	& 4.8\%\\	
	rms time spread (140\,MeV/$c$ settings) [ns]		& 0.40\\
	rms momentum spread (200\,MeV/$c$ settings)	& 4.0\%\\	
	rms time spread (200\,MeV/$c$ settings) [ns]		& 0.34\\
	rms momentum spread (240\,MeV/$c$ settings)	& 3.6\%\\	
	rms time spread (240\,MeV/$c$ settings) [ns]		& 0.31\\
      \hline
    \end{tabular}
  \end{center}
\end{table}
\begin{table}
  \caption{
    Parameters of the cooling-demonstration lattice. 
    $L_{\text{SS} \rightarrow \text{FC}}$ is the distance between the
    centre of the spectrometer solenoid and the centre of the
    neighbouring FC, $L_{\text{FC} \rightarrow \text{FC}}$ the
    distance between the centres of the FCs, and $L_{\text{RF module}
      \rightarrow \text{FC}}$ the distance between the RF module and
    the neighbouring FC. 
  }
  \label{Tab:LatticeParameters}
  \begin{center}
    \begin{tabular}{lcc}
      \hline
      Parameter & Value \\
      \hline
      Length $L_{\text{SS} \rightarrow \text{FC}}$ [mm]       & 2607.5 \\
      Length $L_{\text{FC} \rightarrow \text{FC}}$ [mm]       & 1678.8 \\
      Length $L_{\text{RF module} \rightarrow \text{FC}}$ [mm] & 784.0  \\
      RF Gradient [MV/m]                                 & 10.3   \\
      Number of RF cavities                              & 2      \\
      Number of primary absorbers	                       & 1      \\
      Number of secondary absorbers                      & 2      \\
      \hline
    \end{tabular}
  \end{center}
\end{table}

\begin{figure}
  \begin{center}
    \includegraphics[width=0.9\textwidth]{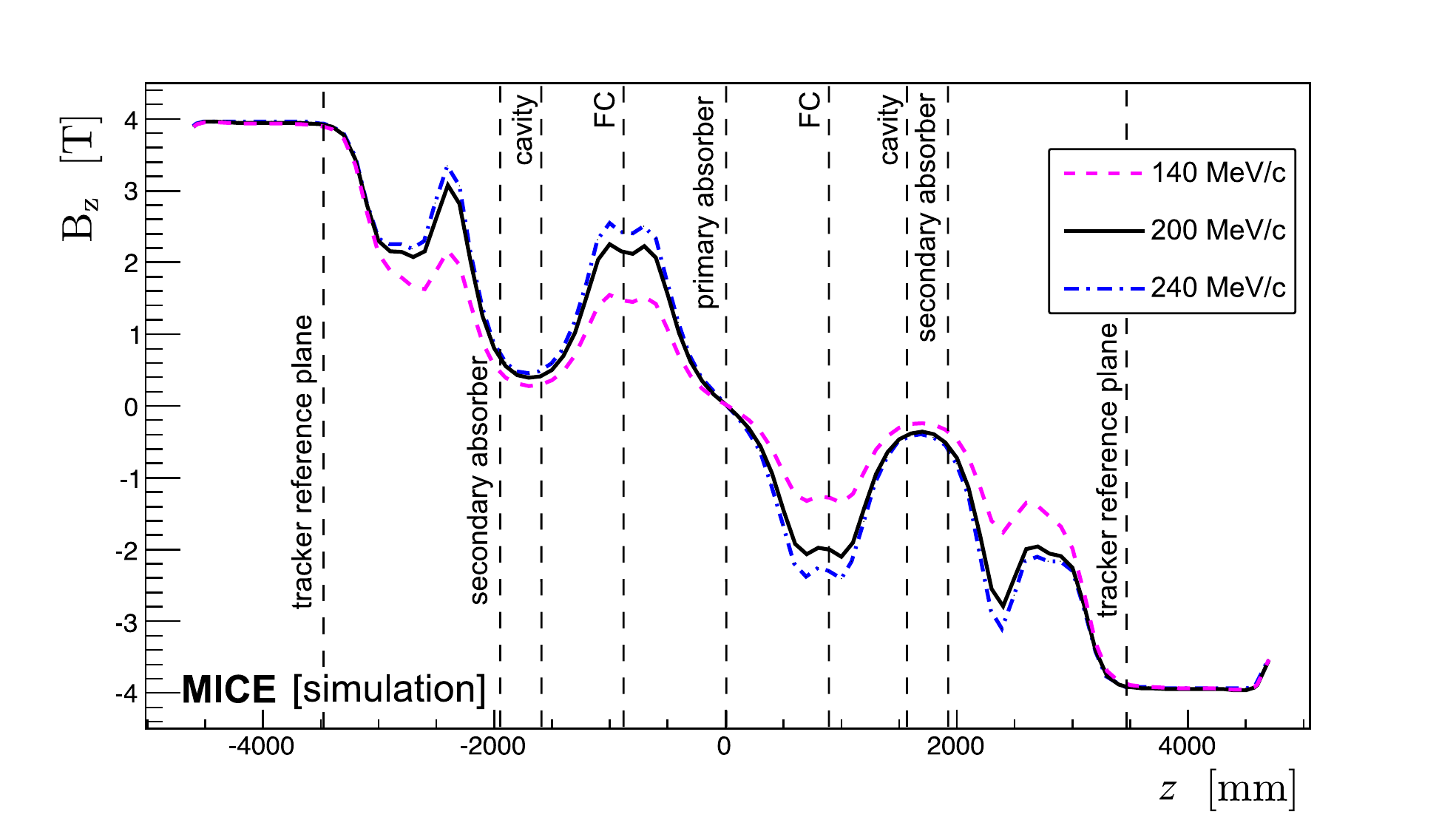}
  \end{center}
  \caption{
    Magnetic field $B_{z}$ on-axis versus the longitudinal coordinate
    $z$ for the cooling-demonstration lattice design for 200\,MeV/$c$
    (solid black line), 140\,MeV/$c$ (dashed purple line) and
    240\,MeV/$c$ (dot-dashed blue line) settings. 
    The vertical dashed lines with labels show the positions of the
    tracker reference planes and the centres of the absorbers, RF
    cavities and focus coil modules. 
  }
  \label{Fig:Bz}
\end{figure}

The currents that produce the optimum magnetic lattice were obtained
using the procedure described above for three momentum settings:
140\,MeV/$c$, 200\,MeV/$c$ and 240\,MeV/$c$.
The magnetic field on axis for each of these settings is shown in
figure \ref{Fig:Bz}.
The fields in the downstream FC and spectrometer are opposite to those
in the upstream FC and spectrometer, the field changing sign at the
primary absorber.
Such a field flip is required in an ionization cooling channel to
reduce the build-up of canonical angular momentum~\cite{PhysRevSTAB.10.064001}. 
The currents required to produce the magnetic fields shown in figure
\ref{Fig:Bz} are listed in table \ref{Tab:CoilCurrents}.
All currents are within the proven limits of operation for the
individual coil windings. 
The magnetic forces acting on the coils have been analysed and were
found to be acceptable.
Configurations in which there is no field flip can also be considered.
\begin{table}
  \caption{
    Coil currents used for 140\,MeV/$c$, 200\,MeV/$c$ and 240\,MeV/$c$
    lattice settings.
  }
  \label{Tab:CoilCurrents}
  \begin{center}
    \begin{tabular}{lccc}
      \hline
	Coil 	& 	140\,MeV/$c$ Lattice [A]	& 200\,MeV/$c$ Lattice [A] & 240\,MeV/$c$ Lattice [A] \\
      \hline
      Upstream  E2		&	$+253.00$	&	$+253.00$	&	$+253.00$	\\
      Upstream C		&	$+274.00$	&	$+274.00$	&	$+274.00$	\\
      Upstream E1		&	$+234.00$	&	$+234.00$	&	$+234.00$	\\
      Upstream M2 		& 	$+126.48$	& 	$+155.37$	& 	$+163.50$	\\
      Upstream M1 		&	$+175.89$	&	$+258.42$	&	$+280.72$	\\
      \hline
      Upstream FC-coil\,1  	&	$+54.14$		&	$+79.35$		&	$+89.77$		\\
      Upstream FC-coil\,2 	& 	$+54.14$		& 	$+79.35$		& 	$+89.77$	 	\\
      \hline
      Downstream FC-coil\,1 	&	$-47.32$ 		&	$-74.10$		&	$-85.35$		\\
      Downstream FC-coil\,2 	&	$-47.32$		&	$-74.10$		&	$-85.35$		\\
      \hline
      Downstream M1 	&	$-140.43$		&	$-231.60$		&	$-261.71$		\\
      Downstream M2 	&	$-100.12$		&	$-149.15$		&	$-159.21$		\\
      Downstream E1 	&	$-234.00$		&	$-234.00$		&	$-234.00$		\\
      Downstream C		&	$-274.00$		&	$-274.00$		&	$-274.00$		\\
      Downstream E2 	&	$-253.00$		&	$-253.00$		&	$-253.00$		\\
      \hline
    \end{tabular}
  \end{center}
\end{table}

\begin{figure}
  \begin{center}
    \includegraphics[width=0.9\textwidth]{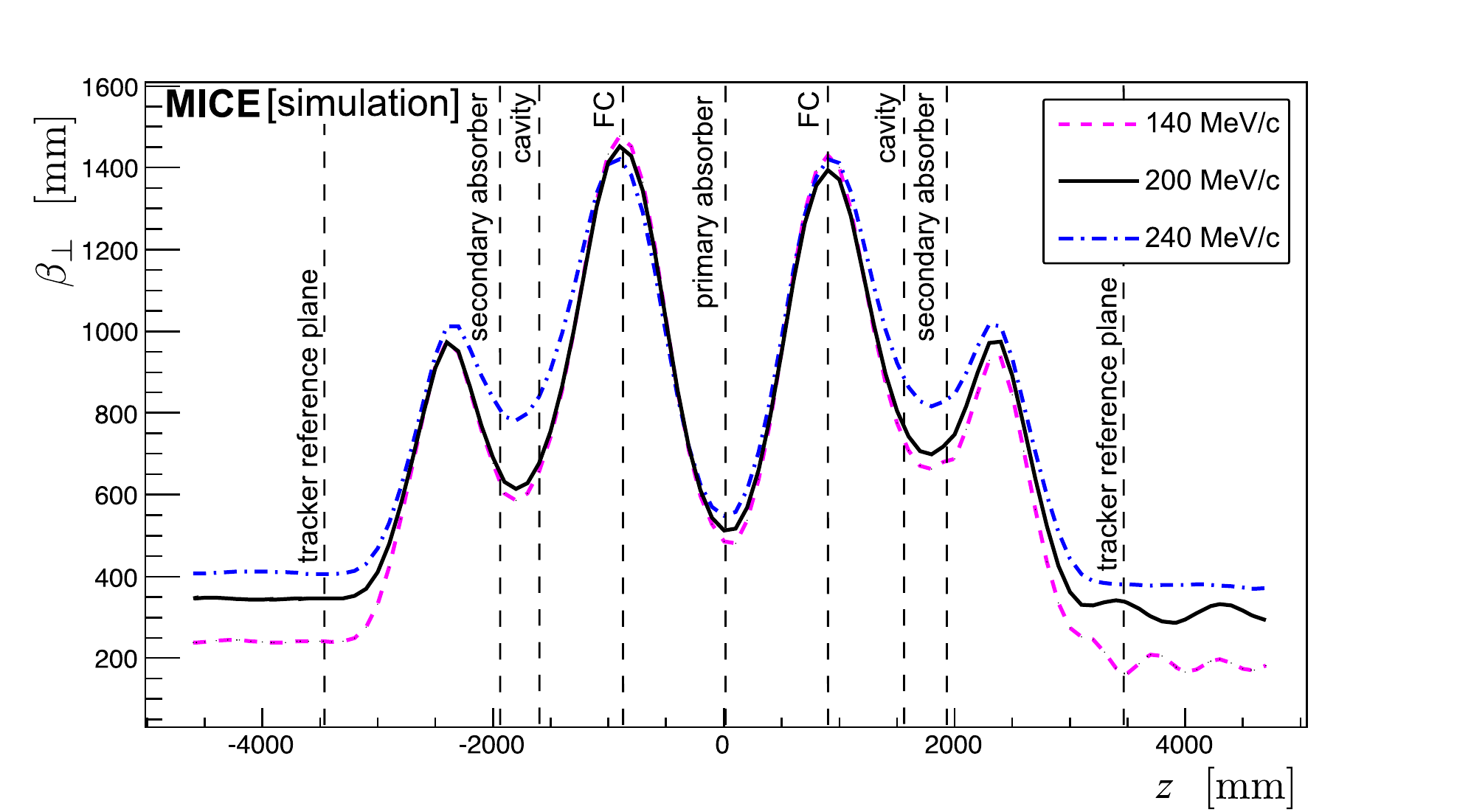}
  \end{center}
  \caption{
    $\beta_{\perp}$ versus the longitudinal coordinate $z$ for
    200\,MeV/$c$ (solid black line), 140\,MeV/$c$ (dashed purple line)
    and 240\,MeV/$c$ (dot-dashed blue line) in the
    cooling-demonstration lattice.
    The vertical dashed lines with labels show the positions of the
    tracker reference planes and the centres of the absorbers, RF
    cavities and focus coil modules.
  }
  \label{Fig:Beta}
\end{figure}

Figure \ref{Fig:Beta} shows matched betatron functions versus
longitudinal position for beams of different initial momentum.
These betatron functions are constrained, within the fiducial-volume
of the trackers, by the requirements on the Courant-Snyder parameters
$\alpha_{\perp}=0$ and $\beta_{\perp}=\frac{2 p_z}{eB_z}$ (where $p_z$
is the mean longitudinal momentum of the beam, $e$ the elementary
charge and $B_z$ the longitudinal component of the magnetic field).
A small betatron-function ``waist'' in the central absorber is
achieved.
Betatron-function values at relevant positions in the different
configurations are summarised in table~\ref{Tab:Beta}.
\begin{table}
  \caption{
    Beta-function values at relevant positions for an initial beam at
    140\,MeV/$c$, 200\,MeV/$c$ and 240\,MeV/$c$ in the
    cooling-demonstration lattice design. 
  }
  \label{Tab:Beta}
  \begin{center}
    \begin{tabular}{lccc}
      \hline
	Parameter & Value for    & Value for    & Value for    \\
	          & 140\,MeV/$c$ & 200\,MeV/$c$ & 240\,MeV/$c$ \\
      \hline
      $\beta_{\perp}$ at primary absorber [mm]  & 480 & 512 & 545 \\
      $\beta_{\perp}$ at upstream secondary absorber [mm] & 660 & 710 & 840	\\
      $\beta_{\perp}$ at downstream secondary absorber [mm] & 680 & 740 & 850	\\
      $\beta_{\perp\mathrm{max}}$ at FC [mm]		&    1480 & 1450 & 1430  \\
      \hline
    \end{tabular}
  \end{center}
\end{table}

\section{Simulation}
\label{Sect:Simu}

Simulations to evaluate the performance of the lattice have been
performed using the official MICE simulation and reconstruction
software MAUS (MICE Analysis User Software) \cite{Tunnell:2011zz}.
In addition to simulation, MAUS also provides a framework for data
analysis. 
MAUS is used for offline analysis and to provide fast real-time
detector reconstruction and data visualisation during MICE running.  
MAUS uses GEANT4 \cite{Ago03,Allison:2006ve} for beam propagation and
the simulation of detector response.
ROOT \cite{Brun:1997pa} is used for data visualisation and for data
storage.

Particle tracking has been performed for several configurations. 
The parameters of the initial beam configurations used for the
simulations are summarised in table \ref{Tab:InitBeam}. 
The simulation of the beam starts at a point between the diffuser and
the first plane of the tracker. 
The beam is generated by a randomising algorithm with a fixed seed.
The number of particles launched for each simulation is a compromise
between the statistical uncertainty required ($\approx 1$\%) and
computing time. 
Each cavity is simulated by a TM$_{010}$ ideal cylindrical pillbox
with a peak effective gradient matched to that expected for the real
cavities. 
The reference particle is used to set the phase of the cavities so
that it is accelerated ``on crest''.
The initial distributions defined in table \ref{Tab:InitBeam} are
centred on the reference particle in both time and momentum.
Table \ref{Tab:Cuts} lists the acceptance criteria applied to all
analyses presented here. 
Trajectories that fail to meet the acceptance criteria are removed
from the analysis.
\begin{table}
  \caption{
    Acceptance criteria for analysis.
  }
  \label{Tab:Cuts}
  \begin{center}
    \begin{tabular}{lcc}
      \hline
      Parameter & & Acceptance condition \\
      \hline
      Particle  & & muon $\mu^+$          \\
      Transmission: pass through two planes & & $z=-4600$\,mm and $z=5000$\,mm\\
      Radius at $z=-4600$\,mm &  & $\leq 150.0$\,mm \\
      Radius at $z=5000$\,mm  &  & $\leq 150.0$\,mm \\
      \hline
    \end{tabular}
  \end{center}
\end{table}

The normalised transverse emittance is calculated by taking the
fourth root of the determinant of the four-dimensional phase-space
covariance matrix \cite{Penn:2000mt,Rogers08beamdynamics}.
The MICE collaboration plans to take data such that the statistical
uncertainty on the relative change in emittance for a particular
setting is 1\%. 
The MICE instrumentation was designed such that the systematic
uncertainty related to the reconstruction of particle trajectories 
would contribute at the $\sim 0.3$\% level to the overall systematic
uncertainty \cite{MICEproposal:2003}; such uncertainties would thus be
negligible.

\graphicspath{{06-Performance/Figures/}}

\section{Performance}
\label{Sect:perf}

\begin{figure}
  \begin{center}
    \includegraphics[width=0.75\textwidth]{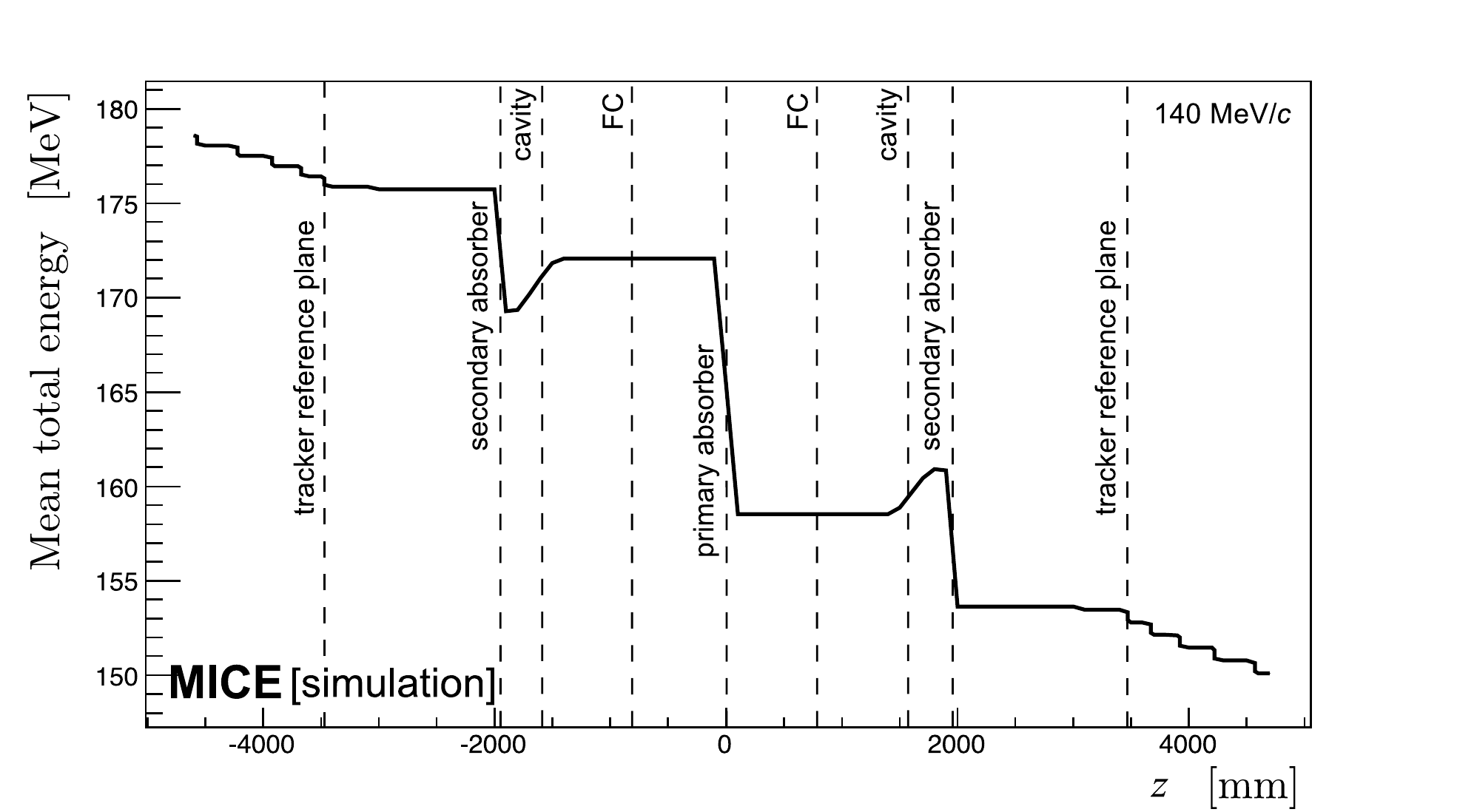}
    \includegraphics[width=0.75\textwidth]{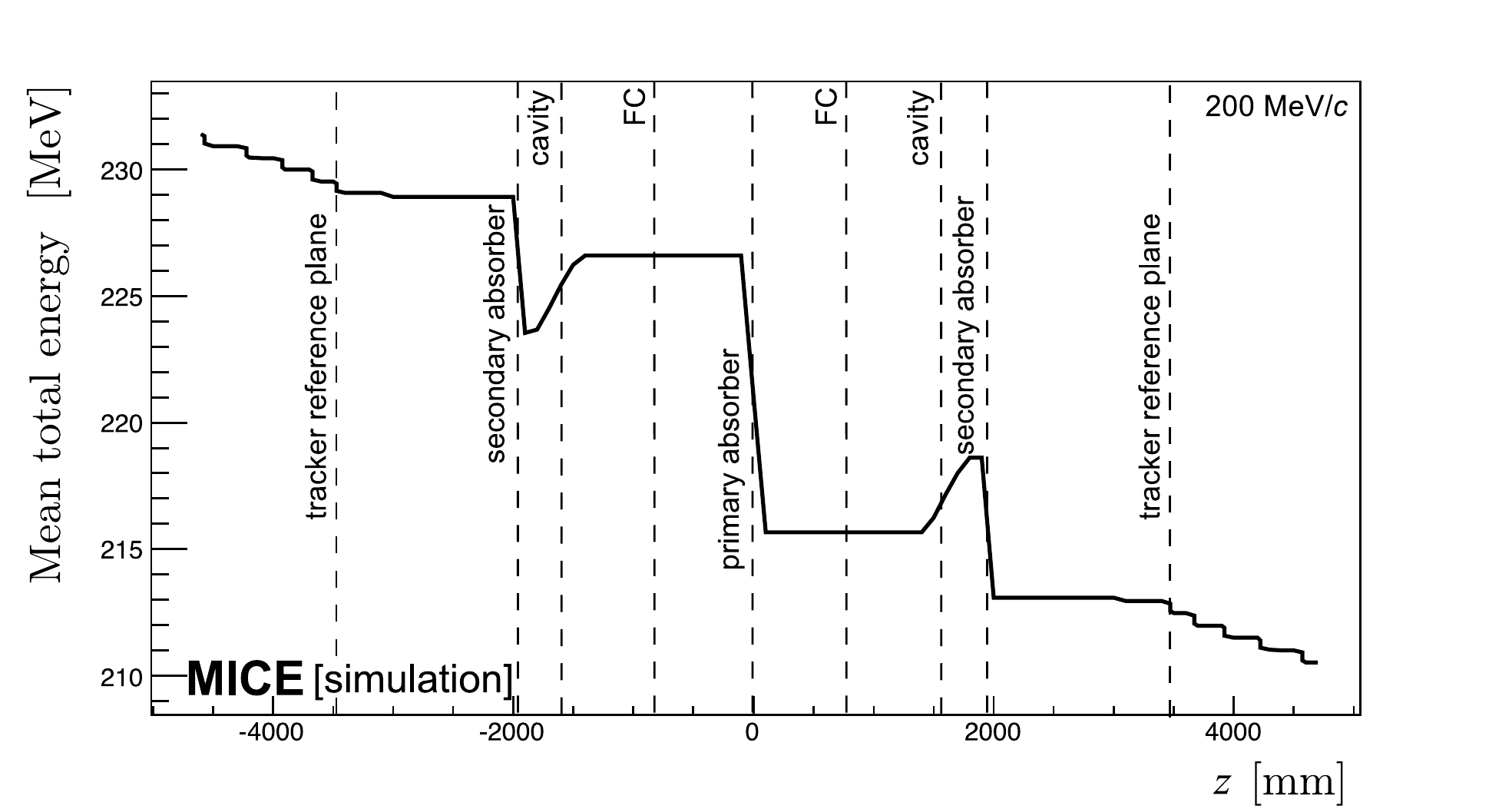}
    \includegraphics[width=0.75\textwidth]{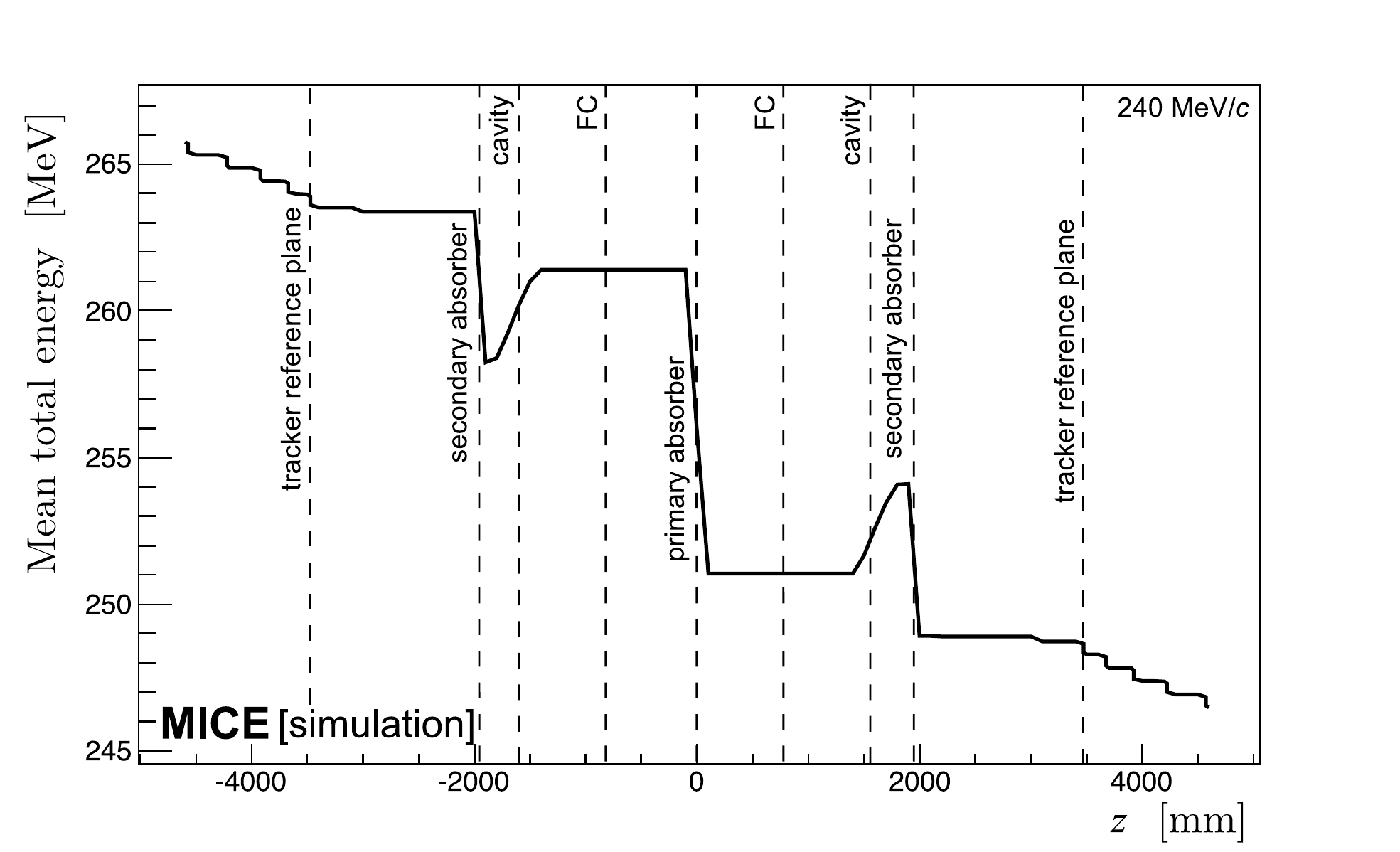}
  \end{center}
  \caption{
    Mean energy of the beam versus longitudinal coordinate
    ($z$) in the cooling-demonstration lattice.
    Top: the 140\,MeV/$c$ configuration for initial emittance
    $\varepsilon_\perp=4.2\,\text{mm}$.
    Middle: the 200\,MeV/$c$ configuration for initial emittance
    $\varepsilon_\perp=6\,\text{mm}$.
    Bottom: the 240\,MeV/$c$ configuration for initial emittance
    $\varepsilon_\perp=7.2\,\text{mm}$.
    The vertical dashed lines with labels show the positions of the
    tracker reference planes, and the centres of the absorbers, RF
    cavities and focus-coil modules. 
  }
  \label{Fig:EnergyAll}
\end{figure}

Figure \ref{Fig:EnergyAll} shows the evolution of the mean energy
of a muon beam as it traverses the lattice.
Beams with initial normalised transverse emittance
$\varepsilon_\perp=4.2\,\text{mm}$, $\varepsilon_\perp=6\,\text{mm}$
and $\varepsilon_\perp=7.2\,\text{mm}$ for initial muon beam momenta
of 140\,MeV/$c$, 200\,MeV/$c$ and 240\,MeV/$c$ respectively are
shown.
The initial normalised transverse emittance is chosen such that the
geometrical emittance of the three beams is the same.
A 200\,MeV/$c$ muon passing through two 32.5\,mm thick secondary LiH
absorbers and one 65\,mm thick primary LiH absorber loses
an energy of 18.9\,MeV. 
Including losses in the scintillating-fibre trackers and windows, this
increases to 24.3\,MeV. 
The accelerating gradient that can be achieved in each of the two
cavities is constrained by  the available RF power and is insufficient
to replace all the lost energy.  
Therefore, a comparison of beam energy with and without acceleration
is required. 
With acceleration an energy deficit of $\langle \Delta E \rangle =
19$\,MeV will be observed.
This measurable difference will be used to extrapolate the measured
cooling effect to that which would pertain if all the lost energy were
restored. 

\begin{figure}
  \begin{center}
    \includegraphics[width=0.7\textwidth]{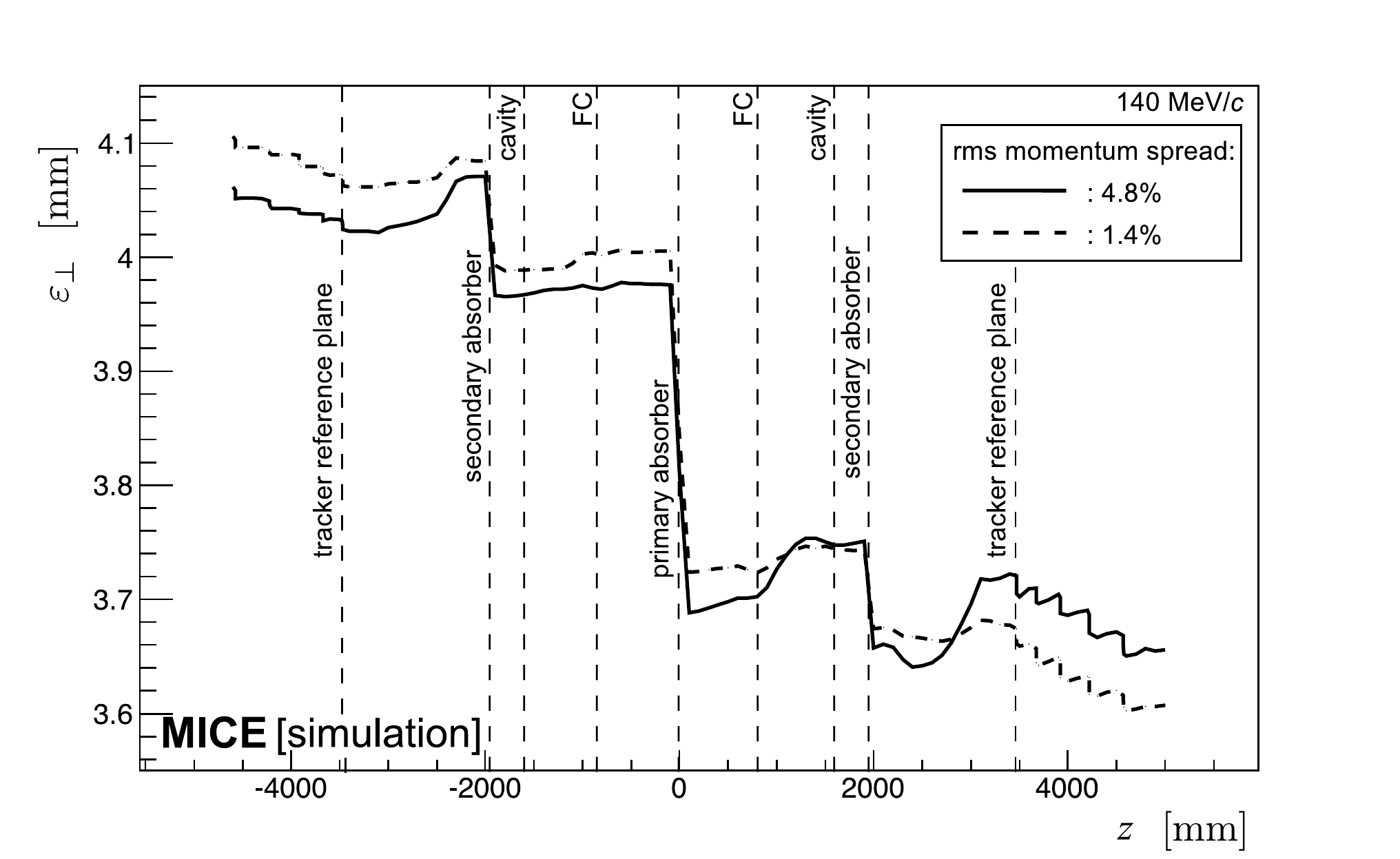}
    \includegraphics[width=0.7\textwidth]{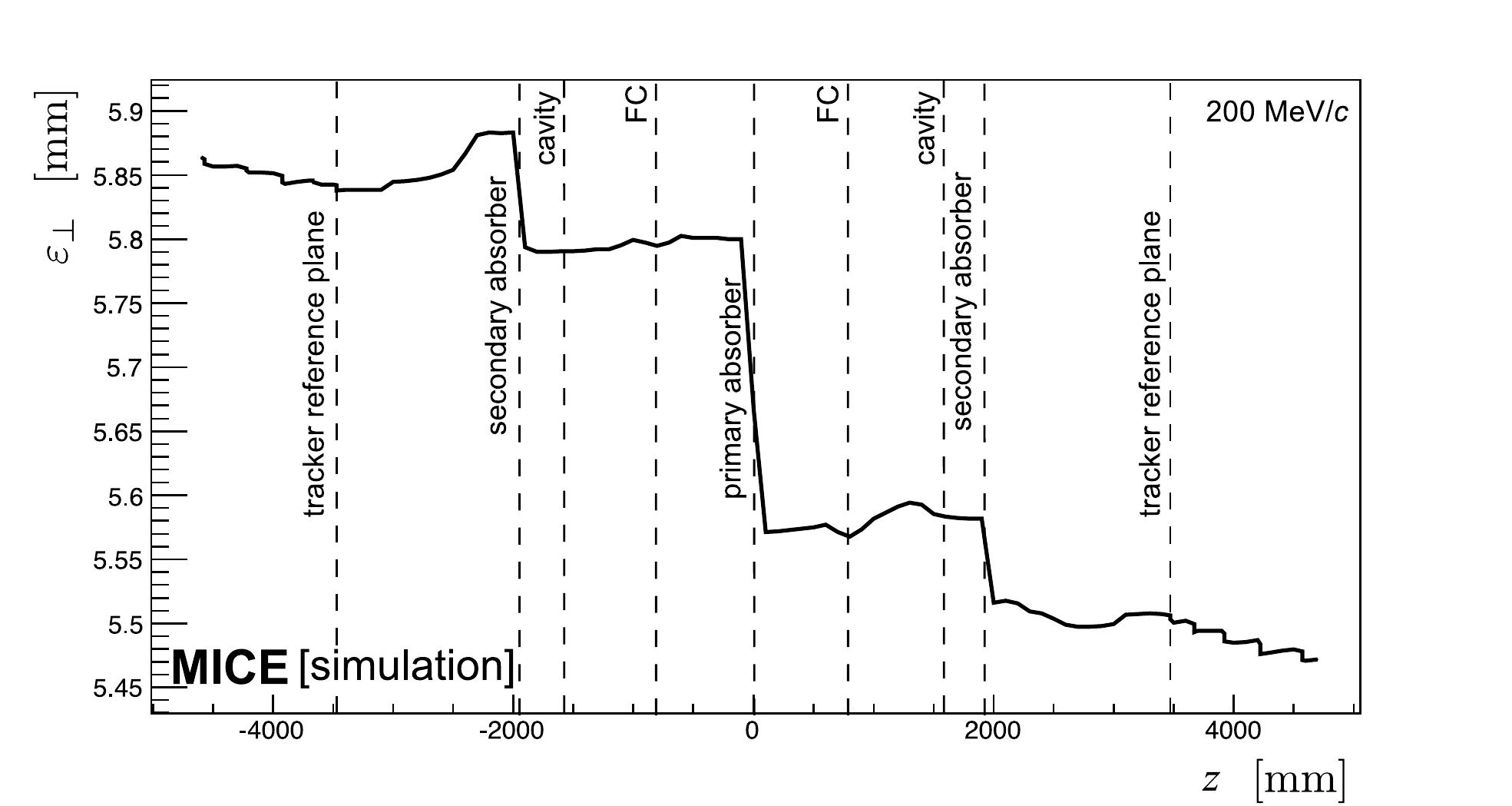}
    \includegraphics[width=0.7\textwidth]{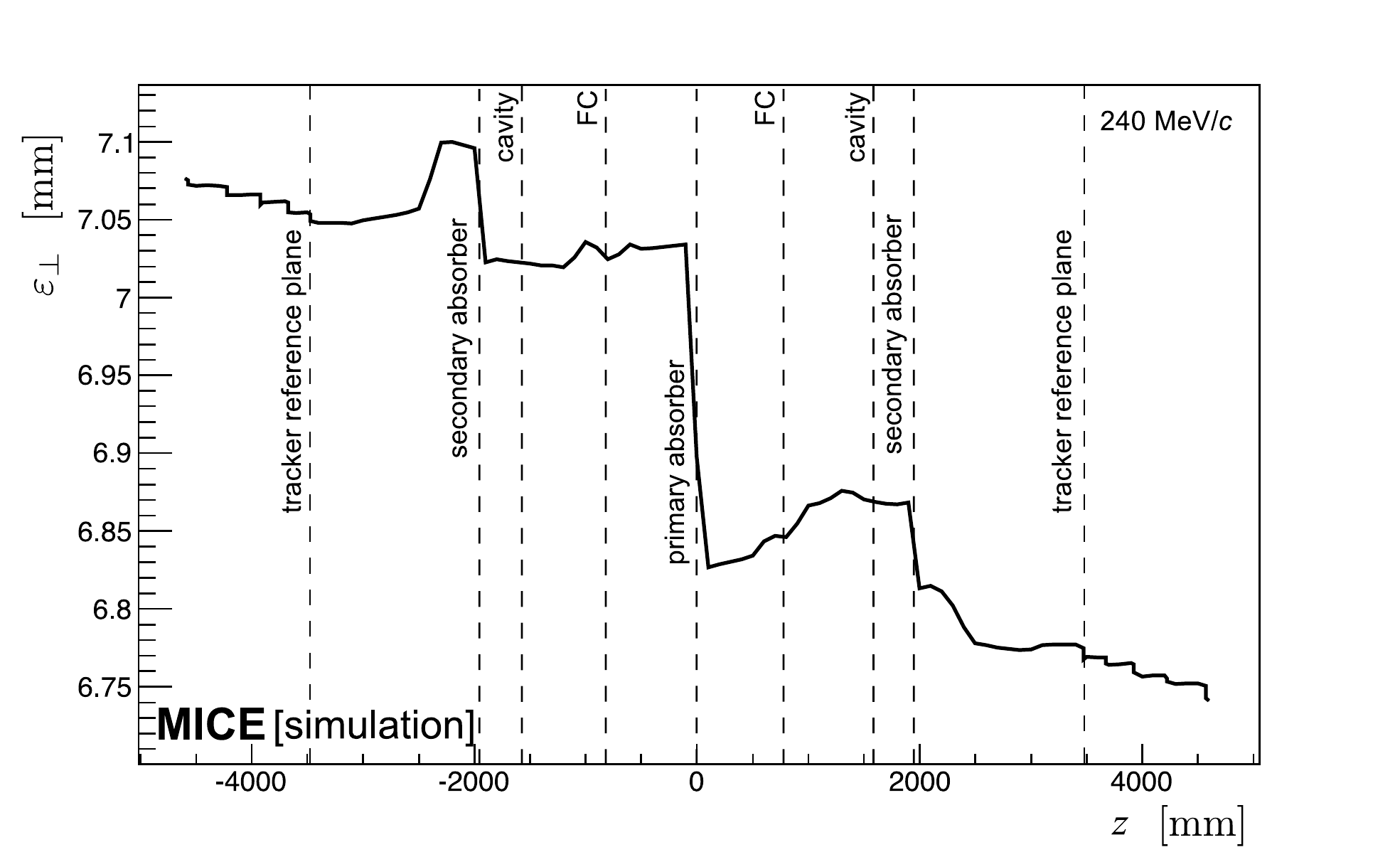}
  \end{center}
  \caption{
    Emittance variation versus the longitudinal coordinate
    ($z$) for the cooling-demonstration lattice design. 
    Top: 140\,MeV/c beam with initial $\varepsilon_\perp=4.2\,\text{mm}$
    with an rms momentum spread of 6.7\,MeV/$c$ (rms spread $4.8\%$,
    solid line) and 2.5\,MeV/$c$ (rms spread $1.8\%$, dashed line).
    Middle: 200\,MeV/c beam with initial
    $\varepsilon_\perp=6\,\text{mm}$ (rms spread $4.0\%$).
    Bottom: 240\,MeV/c beam with initial
    $\varepsilon_\perp=7.2\,\text{mm}$ (rms spread $3.6\%$).
    The vertical dashed lines with labels show the positions of the
    tracker reference planes, and the centres of the absorbers, RF
    cavities and focus coil modules. 
  }
  \label{Fig:Emit200}
\end{figure}

The evolution of normalised transverse emittance across the lattice is
shown in figure \ref{Fig:Emit200}.  
The beam is subject to non-linear effects in regions of high
$\beta_{\perp}$, which cause the normalised transverse emittance to
grow, especially in the 140\,MeV/$c$ configuration. 
This phenomenon can be seen in three different regions of the lattice:
a moderate increase in emittance is observed at $z \approx -2500$\,mm
and $z \approx 1000$\,mm while a larger increase is observed at
$z \approx 3000$\,mm. 
The non-linear effects are mainly chromatic in origin, since they are
greatly lessened when the initial momentum spread is reduced. 
This is illustrated for the 140\,MeV/c case for which the evolution of
normalised emittance for beams with an rms momentum spread of
6.7\,MeV/c and 2.5\,MeV/c are shown.
Nonetheless, in all cases a reduction in emittance is observed between
the upstream and downstream trackers ($z= \pm 3473$\,mm).
The lattice is predicted to achieve an emittance reduction between the
tracker reference planes of $\approx8.1$\%, $\approx5.8$\% and
$\approx4.0$\% in the 140\,MeV/$c$, 200\,MeV/$c$ and 240\,MeV/$c$
cases, respectively. 
A reduction as large as $\approx10$\% can be reached in the
140\,MeV/$c$ configuration with an rms momentum spread of 1.4\%.

\begin{figure}
  \begin{center}
    \includegraphics[width=0.75\textwidth]{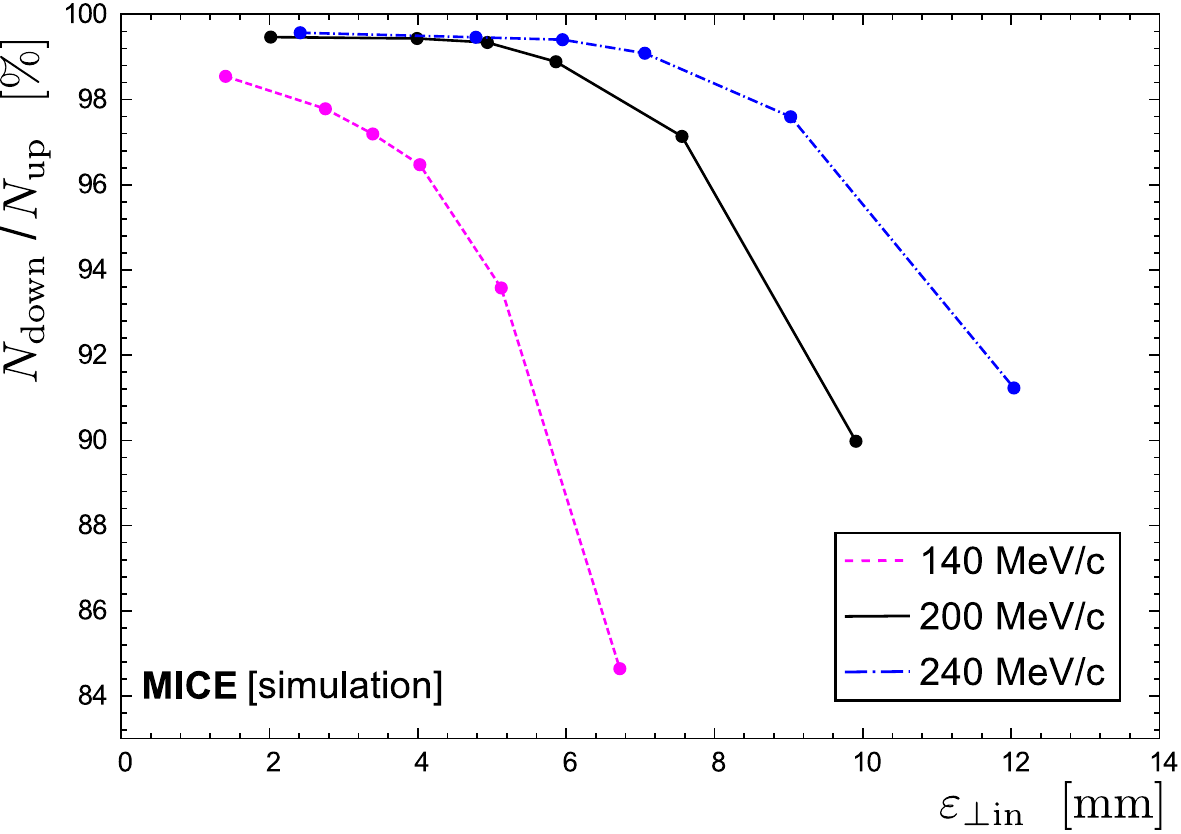}
  \end{center}
  \caption{
    Transmission (defined as the ratio of good muons observed
    downstream of the cooling cell, $N_{\rm down}$, to those observed
    upstream, $N_{\rm up}$) in percent versus initial
    emittance ($\varepsilon_{\perp {\rm in}}$) for the
    cooling-demonstration lattice.
    The transmission of the 140\,MeV/c, 200\,MeV/c and 240\,MeV/c
    lattices are shown as the purple-dashed, solid black, and
    dot-dashed blue lines respectively.
  }
  \label{Fig:Transmission}
\end{figure}

The transmission of the cooling-demonstration lattice for beams of
mean momentum 140\,MeV/c, 200\,MeV/c and 240\,MeV/c is shown in figure
\ref{Fig:Transmission}.
Transmission is computed as the ratio of the number of particles that
satisfy the acceptance criteria observed downstream of the cooling
cell divided by the number that enter the cell.
This accounts for decay losses and implies that, in the absence of
scraping or acceptance losses, the maximum transmission for beams of
mean momentum 140\,MeV/c, 200\,MeV/c and 240\,MeV/c is 98.9\%, 99.2\%
and 99.5\% respectively.
The lattice delivers transmission close to the maximum for 200\,MeV/c
and 240\,MeV/c beams with input emittance below $\approx 5$\,mm and
$\approx 7$\,mm respectively.
For beams of larger input emittance, the transmission gradually
decreases with increasing initial emittance due to the scraping of
high amplitude muons.
The beam is subject to chromatic effects in regions of
high $\beta_{\perp}$, which causes non-linear emittance growth.
The behaviour of the transmission for the various beam energies
results from the different geometrical emittance values of the beam
for the same initial normalised emittance and the energy dependence of
the energy loss and scattering in the material through which the beam
passes.

\begin{figure}
  \begin{center}
    \includegraphics[width=0.75\textwidth]{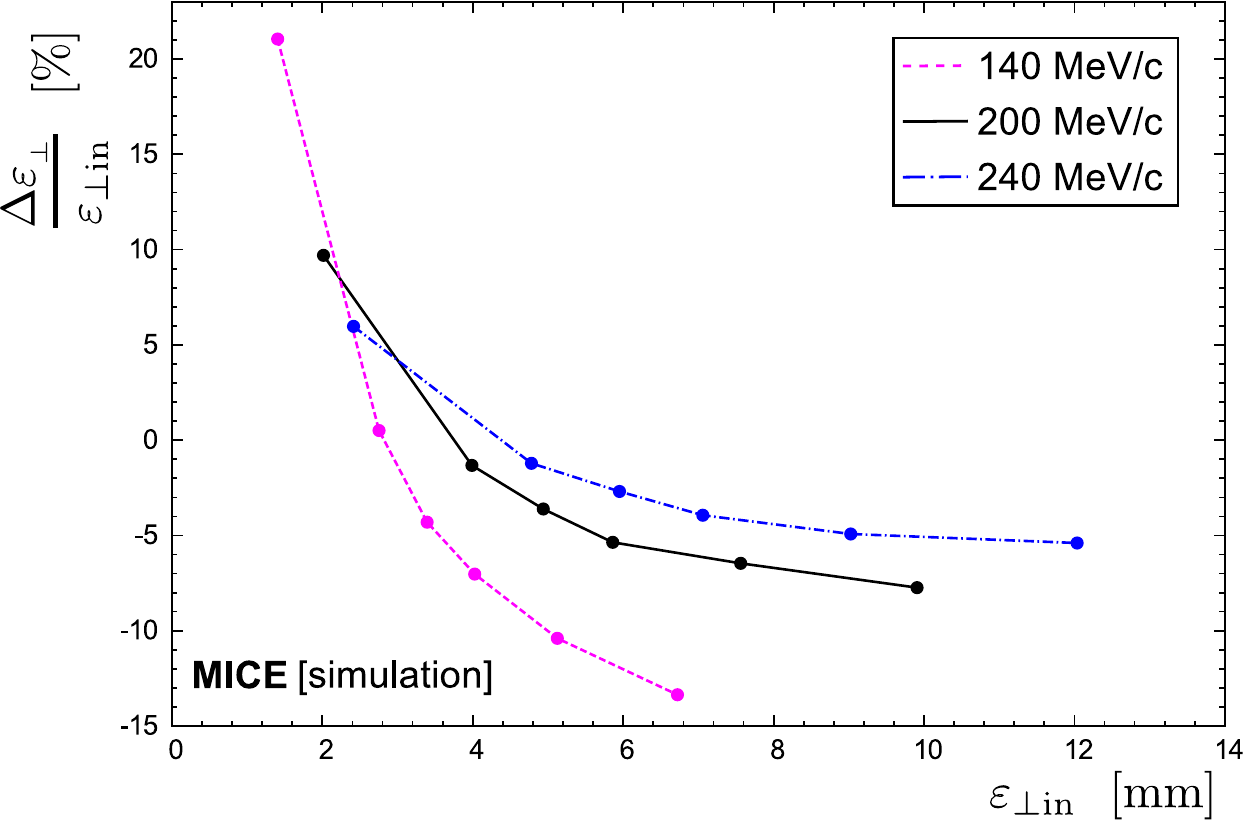}
  \end{center}
  \caption{
    Fractional change in emittance versus initial emittance
    ($\varepsilon_{\perp {\rm in}}$) for the cooling-demonstration
    lattice design measured at the tracker reference planes.
    The fractional change in emittance of the 140\,MeV/c, 200\,MeV/c
    and 240\,MeV/c lattices are shown as the purple-dashed, solid
    black, and dot-dashed blue lines respectively.
  }
  \label{Fig:TransverseEmittanceChangeAll}
\end{figure}

The fractional change in normalised transverse emittance with respect
to the input emittance for beams of mean momentum 140\,MeV/$c$,
200\,MeV/$c$ and 240\,MeV/$c$ is shown in figure 
\ref{Fig:TransverseEmittanceChangeAll}. 
The different values of the equilibrium emittance and the asymptote at
large emittance for each momentum are clearly visible in figure 
\ref{Fig:TransverseEmittanceChangeAll}.
A maximum cooling effect of 15\%, 8\% and 6\% can be observed for
beams with 140\,MeV/$c$, 200\,MeV/$c$ and 240\,MeV/$c$, respectively. 

\section{Conclusion}
\label{Sect:Conclusion}
An experiment by which to demonstrate ionization cooling has been
described that is predicted by simulations to exhibit cooling over a
range of momentum. 
The demonstration is performed using lithium-hydride absorbers and
with acceleration provided by two 201\,MHz cavities.
The equipment necessary to mount the experiment is either in hand
(the superconducting magnets and instrumentation), or at an advanced
stage of preparation.
The configuration of the demonstration of ionization cooling has been
shown to deliver the performance required for the detailed study of
the ionization-cooling technique.

The demonstration of ionization cooling is essential to the future
development of muon-based facilities that would provide the intense,
well characterised low-emittance muon beams required to elucidate
the physics of flavour at a neutrino factory or to deliver multi-TeV
lepton-antilepton collisions at a muon collider. 
The successful completion of the MICE programme would therefore herald
the establishment of a new technique for particle physics.

\section*{Acknowledgements}

The work described here was made possible by grants from Department of
Energy and National Science Foundation (USA), the Instituto Nazionale
di Fisica Nucleare (Italy), the Science and Technology Facilities
Council (UK), the European Community under the European Commission
Framework Programme 7 (AIDA project, grant agreement no. 262025, TIARA
project, grant agreement no. 261905, and EuCARD), the Japan Society
for the Promotion of Science and the Swiss National Science
Foundation, in the framework of the SCOPES programme. 
We gratefully acknowledge all sources of support.
We are grateful to the support given to us by the staff of the STFC
Rutherford Appleton and Daresbury Laboratories.
We acknowledge the use of Grid computing resources deployed and
operated by GridPP in the UK, http://www.gridpp.ac.uk/.

\clearpage
\bibliographystyle{utphys}
\bibliography{Concatenated-bibliography}

\cleardoublepage
\appendix
%
\clearpage
\thispagestyle{plain}
\setlength\parindent{0em}

M.~Bogomilov,  R.~Tsenov, G.~Vankova-Kirilova
\\{\it
   Department of Atomic Physics, St.~Kliment Ohridski University of Sofia, Sofia, Bulgaria
}\\

Y.~Song, J.~Tang
\\{\it
Institute of High Energy Physics, Chinese Academy of Sciences, Beijing, China
}\\

Z.~Li
\\{\it
Sichuan University, China
}\\

R.~Bertoni, M.~Bonesini, F.~Chignoli, R.~Mazza
\\{\it
Sezione INFN Milano Bicocca, Dipartimento di Fisica G.~Occhialini, Milano, Italy
}\\

V.~Palladino
\\{\it
Sezione INFN Napoli and Dipartimento di Fisica, Universit\`{a} Federico II, Complesso Universitario di Monte S.~Angelo, Napoli, Italy
}\\

A.~de Bari, G.~Cecchet
\\{\it 
Sezione INFN Pavia and Dipartimento di Fisica, Pavia, Italy
}
% ...
\\

Y.~Onel
\\{\it
Department of Physics and Astronomy, University of Iowa, Iowa City, IA, USA
}\\

L.~M.~Cremaldi, D.~A.~Sanders, D.~J.~Summers
\\{\it
University of Mississippi, Oxford, MS, USA
}\\

G.~G.~Hanson, C.~Heidt
\\{\it
University of California, Riverside, CA, USA
}\\

\end{document}